\documentstyle[psfig,twocolumn,aps]{revtex}
\begin{document}
\draft
\wideabs{
\title{Polaronic Signatures in Mid-Infrared Spectra: 
Prediction for LaMnO$_3$ and CaMnO$_3$}
\author{Yiing-Rei Chen$^1$, Vasili Perebeinos$^2$ and Philip B. Allen$^1$}
\address{$^1$Department of Physics and Astronomy, State University of
New York, Stony Brook, New York 11794-3800 \\
$^2$Department of Physics, Brookhaven National Laboratory, Upton, New York
11973-5000}
\date{\today}
\maketitle
\begin{abstract}

Hole-doped LaMnO$_3$ and electron-doped CaMnO$_3$ form self-trapped 
electronic states. The spectra of these states have been calculated 
using a two orbital (Mn $e_{\rm g}$ Jahn-Teller) model, from which 
the non-adiabatic optical conductivity spectra are obtained. 
In both cases the optical spectrum contains weight in the gap region, 
whose observation will indicate the self-trapped nature of the carrier 
states. The predicted spectra are proportional to the concentration 
of the doped carriers in the dilute regime, with coefficients 
calculated with no further model parameters. 

\end{abstract}
}

\section{Introduction}

Qualitative discussions of optical absorption from polarons \cite{Emin} have
often been given. For the heavily doped manganites such as
La$_{0.7}$Ca$_{0.3}$MnO$_3$, polarons form at temperatures
$T \geq T_c \approx 250$K where ferromagnetic order is lost. 
Dramatic thermal shifts of optical conductivity 
\cite{Tokura} agree \cite{agree} with the qualitative
polaron picture \cite{qual}.  In the lightly doped
end-member compounds LaMnO$_3$ and CaMnO$_3$, polarons must exist
at low temperature in the spin and orbitally ordered states. 
We have previously \cite{LaMnO3,CaMnO3,PhEmit,STExciton,Raman} 
calculated the properties of these polarons in the 
Mn $e_g$ Jahn-Teller (MEJT) \cite{Millis} model.
Here we point out that the polarons have a rich spectrum of 
local excited states.
We provide a quantitative calculation of the corresponding 
polaron-induced mid-infrared absorption.
The transitions between these local levels are broadened by the
Franck-Condon effect since each local level prefers its own
rearrangement of the lattice.  This is a general phenomenon
expected in many insulators and may account 
for mid-infrared absorption in other systems as well \cite{other}.

In the usual theory of small polaron absorption \cite{Emin,Web},
an electron is trapped at a single site in a non-degenerate orbital.  
The photon permits a transfer of the electron to a first neighbor site.
Direct transfer to the gound state first neighbor polaron (corresponding
to dc conductivity) is exponentially suppressed by unfavorable vibrational
overlap (``Huang-Rhys'') factors.  The largest vibrational overlap is for
transitions into states whose maximum amplitude overlaps with the
zero-point motion of the distorted polaron on the original site.
A Gaussian envoelope of Franck-Condon vibrational sidebands is
predicted.  The cases considered here differ in three ways.
(1) Our electron levels are doubly degenerate before distortion.
(2) Hund's energy is large which blocks hopping unless the
neighbor atom has spin ferromagnetically aligned.
(3) Our polarons are not strictly confined to a single site, and consequently 
have localized excited states in the self-organized potential well.  
These complications are not all unique to manganites, and should
affect polaronic absorption in other materials as well.

Pure LaMnO$_3$ has a single Mn $d$ electron in the $e_g$
doublet ($\psi_2=(x^2 -y^2)/\sqrt{3}, \psi_3=3z^2-r^2$).  Therefore there is a
cooperative Jahn-Teller effect resulting in orbital ordering.
The peak \cite{Torrance,Jung} 
near 2 eV in $\sigma(\omega)$ is interpreted \cite{Jung,STExciton} as
$e_{\rm g} \rightarrow e_{\rm g}$ transitions
across the Jahn-Teller (JT) gap.  Pure 
CaMnO$_3$ has no $e_{\rm g}$ states occupied, and $\sigma(\omega)$ 
peaks at 3.6 eV\cite{Jung} due to excitations across the gap between 
occupied (O $2p$ and Mn $t_{{\rm 2g}\uparrow}$) states 
and empty (Mn $e_{{\rm g}\uparrow}$) states. 
These materials are insulating antiferromagnets (AF) with the nice 
property of being continuously dopable.
Here we focus on two cases, $x=1-\epsilon$ 
(electron-doped CaMnO$_3$) and $x=\epsilon$ (hole-doped LaMnO$_3$.) 
Both materials are insulators when pure, and remain so when lightly
doped because of polaron formation.

We use the model Hamiltonian ${\cal H}_{\rm MEJT}
={\cal H}_t+{\cal H}_{\rm ep}+{\cal H}_{\rm L}+{\cal H}_U$
\cite{LaMnO3}\cite{CaMnO3} similar to the one used by Millis\cite{Millis}, 
except that we have $U=\infty$ for LaMnO$_3$.  For CaMnO$_3$ we include 
also the $t_{\rm 2g}$ AF exchange term ${\cal H}_J$ and do not need the 
Hubbard $U$ term for dilute doping. Two $e_{\rm g}$ orbitals per Mn atom are 
kept to fully represent the symmetries of the orbitals and the 
crystal, and to drive JT distortions.  In the hopping term 
\begin{equation}
{\cal H}_t=\sum_{l,\pm,\alpha\beta}\sum_{\delta=x,y,z}
t_{\alpha\beta}(\pm\hat{\delta})
c_{\alpha}^{\dagger}(\vec{l}\pm\hat{\delta})c_{\beta}(\vec{l}),
\label{eq:hopping}
\end{equation}
there is only one parameter, namely, the 
nearest-neighbor $t=(dd\sigma)$ integral from
Slater-Koster two-center theory. 
In the electron-phonon interaction, we include both the oxygen
breathing vibrations $Q_1$ and the local $E_g$-type oxygen motions
$(Q_2,Q_3)$ which give the Jahn-Teller distortion.
\begin{eqnarray}
{\cal H}_{\rm ep}=-{\rm g}\sum_{\vec{l},\alpha\beta}
c^\dagger_{\alpha}(\vec{l})c_{\beta}(\vec{l})
(Q_3(\vec{l})\sigma^z_{\alpha\beta}&+&Q_2(\vec{l})\sigma^x_{\alpha\beta}
\nonumber \\
&+&\sqrt{2}Q_1(\vec{l})\hat{I}_{\alpha\beta}).
\end{eqnarray}
The indices $\alpha$ and $\beta$ run though orbitals $(\psi_2,\psi_3)$. 
${\cal H}_L$ is the lattice elastic energy of oxygen displacement
along Mn-O-Mn bond directions with an Einstein restoring force.
Note that the JT modes $Q_{\alpha}(\vec{l})$ 
and $Q_{\alpha}(\vec{l}^{\prime})$ are not independent
when Mn sites $\vec{l}$ and $\vec{l}^{\prime}$ are first neighbors,
since the intervening oxygen is
shared by both Mn atoms (this is called the ``cooperative'' phonon effect.)

At the LaMnO$_3$ end of the phase diagram, the dominant term of the 
Hamiltonian after electron-electron repulsion is ${\cal H}_{\rm ep}$.
Instead of delocalizing in a conduction band, holes
in lightly doped LaMnO$_3$ lower their energy more by self-trapping
with local oxygen rearrangement, forming ``anti-JT'' small
polarons\cite{LaMnO3}.
At the CaMnO$_3$ end, at the cost of AF exchange energy, spin-flipping 
in the AF background allows the doped-in electron to lower its energy by 
delocalization.  This spin-polaron effect competes with a localizing
JT polaron effect from ${\cal H}_{\rm ep}$.  The competition
determines the size of the spin-lattice polaron\cite{CaMnO3}.

In this work we calculate the excitation spectrum of the spin-lattice
polaron in CaMnO$_3$ by Born-Oppenheimer approximation, and for LaMnO$_3$
we calculate the excitation spectrum of the self-trapped hole 
in a Cho-Toyozawa \cite{Cho} approximation. 
Our non-adiabatic solutions allow us to 
calculate the Franck-Condon broadened 
optical excitation spectra of these lightly doped 
materials and predict new spectral features that should be 
experimentally visible.

\section{Spin-Lattice Polaron Spectrum in Doped CaMnO$_3$}

Undoped CaMnO$_3$ has Mn$^{4+}$($t_{\rm 2g}^3$) ions with $S=3/2$ 
on an approximately simple cubic lattice, and bipartite (G type) AF 
order below $T_N=125K$. One single doped-in electron will align with the
$S=3/2$ core spin by Hund coupling.  Hopping to first
neighbors is then blocked by the AF order.  To reduce energy by delocalization,
one Mn $S=3/2$ atom flips.  Then the nearby oxygens distort, to
lift the $e_{\rm g}$ degeneracy.  
This gives a seven-site spin-lattice polaron \cite{CaMnO3}. 
The parameters were chosen to be $|t|/JS^2=158$, and $\Gamma=g^2/K|t|=0.25$. 
Each of the seven Mn atoms provides two $e_{\rm g}$ orbitals to form 
a 14-dimensional Hilbert space in which the doped-in electron is 
partially delocalized.

\begin{figure}[htbp]
\centerline{\psfig{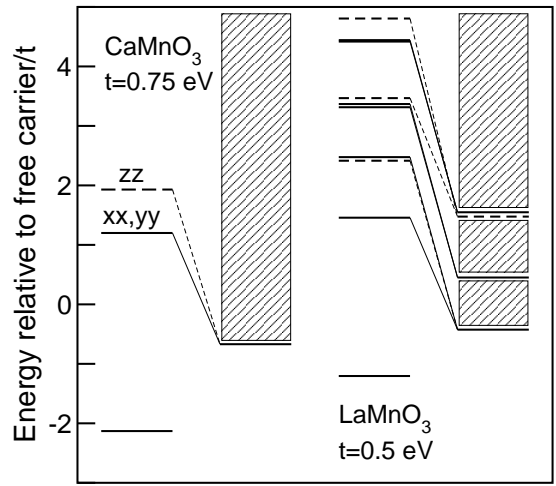}}
\caption{Energy level diagrams for self-trapped states,
relative to the lowest energy of a state in an undistorted environment.
On the left are electron states in CaMnO$_3$ on the right 
are hole states in LaMnO$_3$.  
The lowest level is the ground state of
the doped in $e_g$ electron, bound by spin and lattice distortions.
The dipole-allowed excited levels shown vertically keep the ground
state distortions (dashed lines for $E||\hat{z}$ and solid lines 
for $E\perp\hat{z}$).  Shown to the right are the relaxed energy levels
corresponding to these excited states, and the shaded regions represent 
vibrational excitations accompanying the optical transitions.}
\label{fig:levels}
\end{figure}

The ground state of this polaron is calculated in adiabatic approximation. 
With the local oxygens distorted in the optimized pattern, the ground 
state lies in the gap with energy lowered from the $e_{\rm g}$ band by 
$-2.13 |t|$. The remaining 13 excited states contain three dipole-allowed 
optical excitations. By re-optimizing the oxygen distortion pattern, 
these states (with original energies $1.20 |t|$ and $1.93 |t|$ above 
the $e_{\rm g}$ band) can have their energy lowered to sit below 
the $e_{\rm g}$ band by $-0.67 |t|$. When the electron is optically 
excited to one of these states, oxygens do not have time to 
rearrange to a new optimal pattern, and therefore are left in 
a combination of vibrationally excited states with respect to the 
re-optimized oxygen pattern of the electronic excited state. 
The Franck-Condon principle therefore applies and gives vibrational 
sidebands.  The energy level diagram is shown in Fig. \ref{fig:levels}.

To examine the optical spectrum of this seven-site polaron surrounded 
by 36 oxygens, we start from one of the three symmetry-equivalent 
electronic ground states
$\psi_0(\vec{r},\vec{u_0})$ (e.g., $\theta=0$ \cite{CaMnO3}) 
with local lattice distortion $\vec{u_0}=(u_{0,1},u_{0,2},...,u_{0,36})$
which minimize the energy $E_0(\vec{u})$ of the ground state.
For distortions $\vec{u}$ near $\vec{u_0}$, the vibronic eigenstates associated 
with the electron ground state are approximated by the Born-Oppenheimer 
product:
\begin{equation}
\Psi_{0,n}(\vec{r},\vec{u})=\psi_0(\vec{r},\vec{u_0})\prod_{i=1}^{36}
\phi_{i,n_i}(u_i-u_{0,i}),
\end{equation}
where $\phi_{i,n_i}(u_i-u_{0,i})$ is the $n_i$ phonon state of the
harmonic oscillator centered at $u_{0,i}$.
The usual approximation is made here that $\psi_0(r,\vec{u})$
can be accurately represented by $\psi_0(r,\vec{u_0})$. 
The true electronic ground state varies slowly with
$\vec{u}$, and also depends on the direction 
$\vec{u}-\vec{u_0}$ of deviation from $\vec{u_0}$,
but corrections are small at low temperatures
where the initial state is not vibrationally excited.
Then the adiabatic ground state energy sheet $E_0(\vec{u})$ 
depends quadratically on  $(\vec{u_i}-\vec{u_{0,i}})$ 
and we treat the oxygens as 36 independent 
Einstein oscillators, with energies 
\begin{equation}
E_{0,n}=E_0(\vec{u_0})+\hbar \omega 
\sum_{i=1}^{36}(n_i+\frac 12).
\end{equation}

Among the 13 polaronic excited states, there are only three states that
couple to the ground state by light. They are 
$|\psi_z>\equiv (|\psi_3,\hat{z}>-|\psi_3,-\hat{z}>)/\sqrt{2}$ 
and two other states which are
rotations of $|\psi_z>$ to $\hat{x}$ and $\hat{y}$ 
directions.  Here $|\psi_3,\hat{z}>$ denotes the $\psi_3$ orbital of the
Mn atom sitting immediately above the central Mn. These three excited 
states have their own optimal lattice distortion patterns $\vec{u_z}$, 
$\vec{u_x}$ and $\vec{u_y}$. Starting from $\vec{u_0}$, these electronic 
states remain eigenstates along the paths towards $\vec{u_z}$, 
$\vec{u_x}$ and $\vec{u_y}$ in the lattice distortion space.
Therefore in these chosen directions, the vibronic eigenstates associated
with the electronic excited states have the product form, e.g.:
\begin{equation}
\Psi_{z,n}(\vec{r},\vec{u})=\psi_z(\vec{r},\vec{u_z})\prod_{i=1}^{36}
\phi_{i,n_i}(u_i-u_{z,i}).
\end{equation}

Consider the optical excitation from $\Psi_0$ to $\Psi_z$ for example.
The $T=0$K spectrum is given by
\begin{equation}
\sigma^0(\omega)=\frac{\pi e^2N}{\Omega\hbar}\sum_{n'}
\mid<f,{n'}|\hat{\epsilon}\cdot\vec{p}|i,0>\mid ^2
\delta(\frac{\Delta}{\hbar}+n'\omega_0-\omega)
\end{equation}
The matrix element has both an electronic part and a phonon part. 
The electronic part 
involves the momentum matrix element 
$<\psi_z|p_z|\psi_3,0>$, where $|\psi_3,0>$ denotes the $\psi_3$ 
orbital of the central Mn atom.  This matrix element 
is $<\psi_z|p_z|\psi_3,0>= ima/\hbar<\psi_z|{\cal H}_t|\psi_3,0>$, 
where $m$ is the electron mass 
and $a$ the lattice constant.  This result can be derived by looking at 
FM spin order, and requiring that the corresponding FM Bloch eigenstates 
$|\vec{k}>$ of ${\cal H}_t$ (with energy $\epsilon(\vec{k})$)
should have velocity $<\vec{k}|p_z/m|\vec{k}>=\partial\epsilon(\vec{k})
/\partial\hbar k_z$.

\begin{figure}[htbp]
\centerline{\psfig{figure=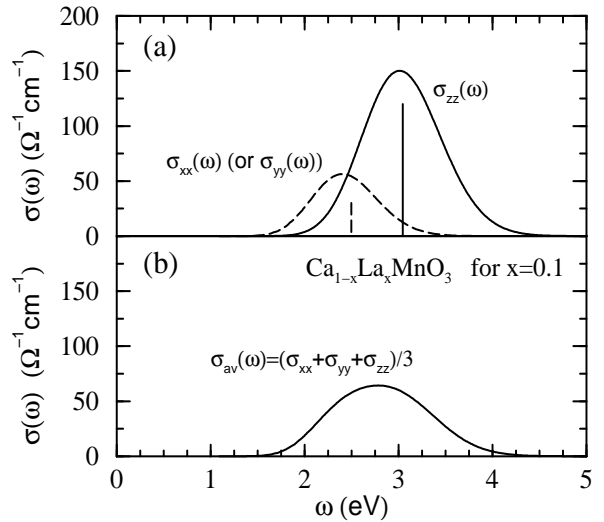,height=2.8in,width=3.0in,angle=0}}
\caption{Polaron-induced optical 
conductivity spectra of lightly doped CaMnO$_3$.
(a) The dashed curve is for $E\perp\hat{z}$ and the solid line for
$E\parallel\hat{z}$. Vertical lines present the adiabatic solution under
fixed oxygen distortions and show the peaks at $2.50$ eV and $3.05$ eV. 
(b) In (a), the solutions belong to one of
the possible ground states ($\theta=0$) \protect\cite{CaMnO3}. There are 
two other possible ground states with orientations $\theta=2\pi/3$, 
and $\theta=4\pi/3$, which permute the directions $\hat{x}$, $\hat{y}$
and $\hat{z}$. Therefore we show the average spectrum. The absolute
value of $\sigma(\omega)$ is shown for 10\% electron doping.}
\label{fig:Cafig}
\end{figure}

The phonon part of the matrix element is 
$\mid<{n'},\vec{u_z}|{n},\vec{u_0}>\mid^2$, where the initial
lattice vibrational state has $n$ vibrational
quanta relative to the oxygen equilibrium positions $\vec{u_0}$. The lattice 
vibrational state is unchanged during the electron excitation, but must
be expanded in the new basis (denoted by quanta ${n'}$) around the new oxygen 
equilibrium positions $\vec{u_z}$.  We consider only zero temperature 
optical conductivity, which restricts the initial vibrational 
quantum numbers ${n}$ to be zero.  The non-zero overlaps of this initial 
phonon state with the new phonon basis states of the displaced lattice
give a spectrum with a Gaussian profile of delta function peaks of the 
multi-phonon transitions accompanying the electron exitation.  The 
phonon frequency is taken to be $\hbar\omega=0.084$ eV. 
This is a little higher than the stretching frequency
reported from Raman data \cite{vibr}, but the discrepancy does
not affect our conclusions. 
The hopping integral $t$ is taken from density-functional
calculations \cite{Pickett} to be $t=-0.75$ eV. The delta 
functions are broadened into Gaussians of width $\sqrt{n'}\gamma$
where $\gamma=0.014 eV$ and $n'$ is the total vibrational quantum.
The results are shown in Fig. \ref{fig:Cafig}.

\section{Polaronic Spectra in Hole Doped LaMnO$_3$}

In LaMnO$_3$ the electron-phonon term ${\cal H}_{\rm ep}$ stabilizes an
orbitally ordered ground state {\sl via} a cooperative JT oxygen distortion.
We use the same model as in CaMnO$_3$, with different filling, and fit it 
to the JT ground state. We previously used this model to predict the
resonant multi-phonon Raman cross-section \cite{Raman} and the angle 
resolved photoemission spectrum including the Franck-Condon broadening
\cite{PhEmit}.
The photoemission process creates a sudden hole without time for oxygen 
relaxation. Thus it is in a vibrationally excited state.
In the doped compound a hole is initially present    
in its ground vibrational state in a locally distorted oxygen environment.
The zeroth order electronic many-body wave function has a hole $h$ in 
the orbital at the origin $0$ (taken arbitrarily to be an $x$-oriented
orbital) in an otherwise perfectly correlated state with one $E_g$
electron per atom.
\begin{eqnarray}
|h0>&=&c_X(0)|{\rm JT}>
\nonumber \\
|{\rm JT}>&=&\prod_{\ell}^A c_X^{\dagger}(\ell)
    \prod_{\ell^{\prime}}^B c_Y^{\dagger}(\ell^{\prime})|0>
\label{gs}
\end{eqnarray}
Here the orbitals $\psi_X$ and $\psi_Y$ are orthogonal $x-$ and
$y-$oriented orbitals $(\psi_2 \pm \psi_3)/\sqrt{2}$.

At the next level of approximation the state $|h0>$ is coupled to 18 
other states.    Six of these states keep the hole at the origin
but have an orbital defect, or ``orbiton'' $o$ on one of the surrounding
Mn sites $\pm\hat{\ell}$ where $\hat{\ell}$ is $\hat{x}$, $\hat{y}$,
or $\hat{z}$,
\begin{equation}
|h0,o\pm\hat{\ell}>=c_X^{\dagger}(\pm \hat{\ell})c_Y(\pm \hat{\ell})|h0>
\label{oexc}
\end{equation}
Six more states have the hole moved to the neighboring Mn site,
with no orbital defect,
\begin{equation}
|h\pm\hat{\ell}>=c_Y(\pm \hat{\ell})c_X^{\dagger}(0)|h0>.
\label{hexc1}
\end{equation}
The last six states have the hole on a neighbor and leave a
misoriented $|Y>$ orbital at the origin,
\begin{equation}
|h\pm\hat{\ell},o0>=c_Y(\pm \hat{\ell})c_Y^{\dagger}(0)|h0>.
\label{hexc2}
\end{equation}

The Hamiltonian ${\cal H}_{\rm MEJT}$ becomes a $19\times19$ 
matrix in this truncated Hilbert space.

In the adiabatic solution of the zeroth order hole
ground state $|h0>$, the neighboring oxygens in 
$\pm \hat{x}, \pm \hat{y}, \pm \hat{z}$ directions are distorted to 
positions given in Table \ref{tab}, row 1, which minimize the polaron 
energy \cite{LaMnO3}.  

With the oxygen pattern frozen in these positions, the diagonal element 
$<h0|{\cal H}_{\rm ep}|h0>$ is set to zero.  Non-zero $Q_3$ distortions on 
the neighboring sites couple the state $|h0>$ to the six orbital excitations
$|h0,o\pm\hat{\ell}>$.
Row 2 of Table \ref{tab} gives the diagonal elements 
$<h0,o\pm\hat{\ell}|{\cal H}_{\rm ep}|h0,o\pm\hat{\ell}>$
and row 3 gives the corresponding off-diagonal elements. 
The electron hopping term ${\cal H}_t$ couples the state $|h0>$
to states with the hole moved to the one of the six neighboring sites,
both without (Eq. \ref{hexc1}) and with (Eq. \ref{hexc2}) an
orbital defect.  The energies of these states
are given in Table \ref{tab} rows 4 and 5.

\begin{figure}
\psfig{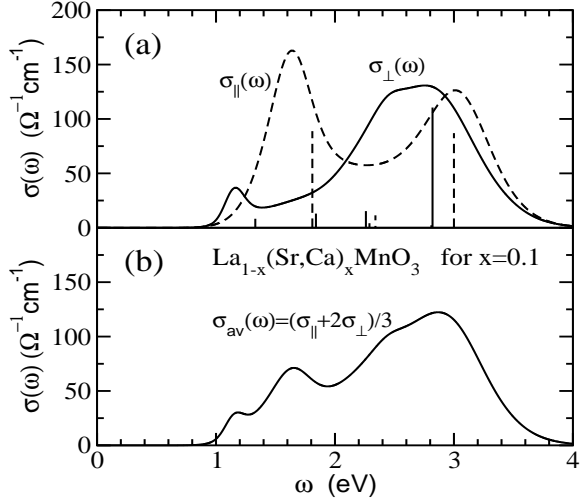}
\caption{Polaron-induced optical conductivity 
spectra of lightly doped LaMnO$_3$.
(a) The dashed line is for $E \parallel \hat{z}$ and solid line for
$E \perp \hat{z}$. Vertical lines is an adiabatic solution (frozen
oxygen distortions). The non adiabatic spectra consist of three and six
Franck-Condon broadened peaks for $E \parallel \hat{z}$ and
$E \perp \hat{z}$ respectively. (b) Shows the average over polarizations.
The overall spectral weight is proportional to the hole concentration and is
given for 10\% doping.}  
\label{Lafig}
\end{figure}

Diagonalization of this 19$\times$19 matrix gives the ground state polaron
solution and the excited spectra.
Among the 18 excited states, there are 9 odd states which are coupled by
light to the even parity ground state with momentum matrix elements
$<\Psi|\hat{p}|h0>=ima/\hbar<\Psi|{\cal H}_t|h0>$, whereas the other 9 even
states are optically silent.
No new parameter is needed for dipole matrix elements to predict the
absolute value of $\sigma(\omega)$.
The spectra would consist of 9 delta functions, three for $E$-field
polarized in $\hat{z}$ direction $\sigma_{\perp}(\omega)$ and six for
$\sigma_{\parallel}(\omega)$ shown by vertical lines
on Fig. (\ref{Lafig}, a).
The parameters were chosen to be $\Delta=0.95$ eV,
$t=0.5$ eV, $\hbar\omega=0.075$ eV, and Mn-Mn distance $a=4.0\AA$.


In order to investigate the lattice-coupled electronic
excitation, the surrounding oxygens should be allowed to vibrate.
Since the frozen oxygen positions are not optimal for the excited
hole states, Franck-Condon vibrational sidebands should appear in the
optical spectrum.  Assuming that the individual vibrational sidebands are
not energy-resolved, there will be broadening by an amount roughly equal to
the energy gain due to structural relaxation.  In order to quantitatively
describe the process, a theory beyond the adiabatic approximation is needed.   

We use an approach similar to that applied to describe photoemission
in the undoped material \cite{PhEmit}.
The method was proposed by Cho   and Toyazawa \cite{Cho}, who used an
infinitely large truncated Hilbert space to diagonalize a 1D polaronic
Hamiltonian.  
For each of the states appearing in Eq. (\ref{oexc}), Eq. (\ref{hexc1}) 
and Eq. (\ref{hexc2}), we re-optimize the positions of the 6 oxygens 
surrounding the hole. The relaxed energies are listed in Table \ref{tab}, 
row 6, 7, and 8.  Notice that for each of the states in Eq. (\ref{oexc}), 
the energy gain from relaxation is $\Delta$, similar to exciton creation 
in the parent compound \cite{STExciton}. 
Each of these states, including $|h0>$, is now allowed to have arbitrary 
number of vibrational quanta of the surrounding oxygens around their 
re-optimized positions for that particular state.  This adds additional
vibrational multiples $n\hbar\omega$ to the relaxed energy.
In this chosen basis, the ${\cal H}_{\rm ep}+{\cal H}_{\rm L}$ term of the
Hamiltonian is diagonal. The complete Hamiltonian with the off-diagonal
term ${\cal H}_{\rm t}$ can be diagonalized exactly. The hole ground state
is not vibrationally excited at zero temperature, and the electronic part
of this ground state has even symmetry which is coupled by light to odd 
excited states.

The Cho-Toyozawa calculation can be simplified by ignoring the sharing of the 
oxygens by its two nearest Mn neighbors.  This partial
removal of the ``cooperative phonon'' constraints was found
in our photoemission calculation \cite{PhEmit} to make only a
small error.
The energies of the odd parity states can be obtained as roots of the
following analytic equation:
\begin{eqnarray}
&&G_{\ell}^1(E)G_{\ell}^2(E)+G_{\ell}^1(E)G_{\ell}^3(E)=1
\nonumber\\
&&G_{\ell}^i(E)=t_{\ell}^i\exp{(-a_{\ell}^i)}\sum_{n=0}^{\infty}
\frac{1}{E-b_{\ell}^i-n}\frac{(a_{\ell}^i)^n}{n!}
\label{solut}
\end{eqnarray}
The roots $E$ give energies of the excited states
for light polarized in $\ell=\hat{x},\hat{y},\hat{z}$ directions.
The parameters $a_{\ell}^i$, $b_{\ell}^i$ and $t_{\ell}^i$ in 
Eq. \ref{solut} are given in table \ref{tab}.
The meaning of parameters $a_{\ell}^i$ is the energy gain
due to the structural rearrangement. 
The parameters $b_{\alpha}^i$ are energies of the
electronic excited states in vibrational ground state with relaxed oxygen  
distortions. 
The hopping matrix elements $t_{\alpha}^i$ couple the ground state to the 
electronically excited states and they are proportional to the dipole matrix 
elements of the allowed transitions. 
The parameters $\Delta$ and $t$ used in Eq. \ref{solut} are all in units of
$\hbar\omega=75$ meV.

The energy of the ground state is lowered by 0.39 eV due to the hole  
delocalization
on the neighboring sites such that only 83 \% of the polaron is present on 
the
central site. This value is very close to the estimation by second order
perturbation theory \cite{LaMnO3}, which gives
$-0.167t/\Gamma-0.488\Gamma t=0.41$ eV ($\Gamma=\Delta/(8t)$).
The excitation spectrum is measured from the
renormalized ground state energy and shown on Fig. \ref{Lafig}. It should be
noted that the electronically excited states have their maximum weight on the
neighboring sites and are not allowed to lower their energies by
spreading on the next neighboring sites due to the truncation of the
Hilbert space. We have done an estimate of the finite Hilbert space effect
by comparing the adiabatic spectra using 19 electronic states on 7 sites 
with those from the 79 states on 27 sites including the next neighboring 
atoms. The spectrum in a larger basis has been shifted by approximately 
0.3 eV to
lower energies. The energy peak positions and relative intensities
in the adiabatic solution follow the trend of the non-adiabatic result
(see Fig \ref{Lafig}.a). We expect an overall shift of the spectrum in
Fig. \ref{Lafig} to the lower energy by about 0.3 eV, such that the
the spectrum sets in at about 0.7 eV.

%


\section{Conclusion}

The particular details
for the manganites show that interesting processes occur
which were not fully anticipated in the qualitative discussions
previously given \cite{Emin}.  Similar processes should be at work in
other insulators such as cuprates and nickelates. 

New spectral features are predicted to
appear in the JT gap of lightly doped LaMnO$_3$, and in the
insulating gap of lightly doped CaMnO$_3$.  The excitation spectra 
are calculated in non-adiabatic Born-Oppenheimer or Cho-Toyozawa 
approximation.
These excitations are dipole-allowed with dipole matrix elements 
proportional to the hopping parameter $t$. 

In the high $T$ paramagnetic phase, in a mean field approximation,
spin disorder has the effect of reducing the parameter $t$ 
by 30\%.  This estimate suggests that the spectral weight 
due to hole excitations in LaMnO$_3$, for example, is reduced by 50\% 
above the N\'eel temperature $ T_N\approx 140K$ \cite{Wollan}, 
whereas the orbiton excitation centered 
at 2 eV \cite{STExciton} is not sensitive to magnetic order and hence 
is temperature independent.  In CaMnO$_3$, above $T_N\approx 125K$ 
\cite{Wollan},
spin polarons should persist in disordered and increasingly delocalized
form.  This will cause the spectrum of Fig. \ref{fig:Cafig} to broaden
and shift to lower energy.

Our picture makes the prediction that,
at frequencies well below the lowest weak peak in the sub-gap region,
there should be a weak Urbach tail with characteristic Urbach
temperature dependence.  The tail is from the lowest-energy processes,
and are weak because of small vibrational overlap.
The available experimental data  \cite{Jung} on doped compounds 
agree with our picture.
Further experiments on more lightly-doped materials, and
especially transmission experiments through thin single
crystals, would give a better test of our theory.

\acknowledgments
This work was supported in part by NSF Grant No.\ DMR-0089492 and
by DOE  Grant No.\ DE-AC-02-98CH10886.

\pagebreak
\begin{table}
\caption{The parameters of the single-hole version of Hamiltonian 
${\cal H}_{\rm MEJT}$ for oxygen displacements (row $1$), 
the diagonal elements 
$<h0,o\pm\hat{\ell}|H|h0,o\pm\hat{\ell}>$ are shown in row $2$, 
the off-diagonal coupling elements $gQ_3({\ell})$ are in row $3$, 
the diagonal elements $<h\pm\hat{\ell}|H|h\pm\hat{\ell}>$ are in row $4$,
the diagonal elements $<h\pm\hat{\ell},o0>|H|h\pm\hat{\ell},o0>$ 
are in row $5$. The relaxed 
energies of each of the states $|h0,o\pm\hat{\ell}>$ are in row $6$, 
$|h\pm\hat{\ell}>$ are in row $7$ and $|h\pm\hat{\ell},o0>$ are in row $8$.
The parameters $a_{\ell}^i$, $b_{\ell}^i$ and $t_{\ell}^i$ ($i=1,2,3$) 
used in Eq. \protect{\ref{solut}} are given in the rest of the table.}
\label{tab}
\begin{tabular}{c|ccc}
 $\hat{\ell}=$ & $\pm\hat{x}$ &  $\pm\hat{y}$ &  $\pm\hat{z}$ \\
\hline
$1$ & $\pm (\sqrt{4/3}-1)g/K$ & $\mp (\sqrt{4/3}+1)g/K$ & 
$\mp \sqrt{4/3}g/K$ \\
$2$ & $(7-2/\sqrt{3})\Delta/4$  & 
$(7+2/\sqrt{3})\Delta/4$ & $2\Delta$ \\
$3$ & $(2\sqrt{2}+\sqrt{6})\Delta/24$ & $(2\sqrt{2}-\sqrt{6})\Delta/24$  
& $-\sqrt{2}\Delta/6$  \\
$4$ &  $37\Delta/24$ & $37\Delta/24$ & $40\Delta/24$ \\
$5$ &  $61\Delta/24$ & $61\Delta/24$ & $64\Delta/24$ \\
$6$ & $(3-2/\sqrt{3})\Delta/4$ & $(3+2/\sqrt{3})\Delta/4$ & 
$\Delta$ \\
$7$ & $0$ & $0$ & $0$ \\
$8$ & $(3+2/\sqrt{3})\Delta/4$ & $(3-2/\sqrt{3})\Delta/4$ & $\Delta$ \\
$a^1_{\ell}$ & $\Delta$  & $\Delta$  &  $\Delta$ \\ 
$a^2_{\ell}$ & $37\Delta/24$  & $37\Delta/24$  &  $40\Delta/24$ \\
$a^3_{\ell}$ & $(43/3-4/\sqrt{3})\Delta/8$  & 
$(43/3+4/\sqrt{3})\Delta/8$  &  $40\Delta/24$ \\
$b^1_{\ell}$ & $(6-4/\sqrt{3})\Delta/8$  & $(6+4/\sqrt{3})\Delta/8$ &$\Delta$\\
$b^2_{\ell}$ & $0$  & $0$ & $0$ \\
$b^3_{\ell}$ & $(6+4/\sqrt{3})\Delta/8$  & $(6-4/\sqrt{3})\Delta/8$ &$\Delta$\\
$t^1_{\ell}$ & $(2+\sqrt{3})t/4$  & $(2-\sqrt{3})t/4$  & $t/2$ \\
$t^2_{\ell}$ & $(2+\sqrt{3})t/4$  & $(2-\sqrt{3})t/4$  & $t/2$ \\
$t^3_{\ell}$ & $(2-\sqrt{3})t/4$  & $(2+\sqrt{3})t/4$  & $t/2$ \\
\end{tabular}
\end{table}

\end{document}